# SoftCloud: A Tool for Visualizing Software Artifacts as Tag Clouds

Ra'Fat Al-Msie'deen[*]

Department of Computer Information Systems, Faculty of IT
Mutah University, P.O. Box 7, Mutah 61710, Karak, Jordan
E-mail: rafatalmsiedeen@mutah.edu.jo

**Abstract**—Software artifacts visualization helps software developers to manage the size and complexity of the software system. The tag cloud technique visualizes tags within the cloud according to their frequencies in software artifacts. A font size of a tag within the cloud indicates its frequency within a software artifact, while the color of a tag within the cloud uses just for aesthetic purposes. This paper suggests a new approach (SoftCloud) to visualize software artifacts as a tag cloud. The originality of SoftCloud is visualizing all the artifacts available to the software program as a tag cloud. Experiments have conducted on different software artifacts to validate SoftCloud and demonstrate its strengths. The results showed the ability of SoftCloud to correctly retrieve all tags and their frequencies from available software artifacts.

أداة لتصور وثائق البرنامج على شكل سحابة العلامات

رافت احمد المسيعدين

قسم نظم المعلومات الحاسوبية - كلية تكنولوجيا المعلومات
جامعة مؤتة، الكرك - الأردن - الرمز البريدي (61710)
البريد الإلكتروني: rafatalmsiedeen@mutah.edu.jo

**الملخص:** يساعد تصور وثائق البرنامج (software artifacts) مطوري البرامج على إدارة حجم وتعقيد البرنامج. تقنيات التصور المستندة إلى سحابة العلامة (tag cloud)، تصور العلامات داخل السحابة وفقا لمعدل تكرارها في وثائق البرنامج. يشير حجم خط العلامة (font size) داخل السحابة إلى تردد العلامة في وثيقة البرنامج. يستخدم لون العلامة (color) داخل السحابة لأغراض جمالية فقط. يقترح هذا البحث نهجًا جديدا (SoftCloud) لتصور وثائق البرنامج على شكل سحابة العلامات. تكمن أصالة SoftCloud في أنها تصور جميع الوثائق المتاحة للبرنامج على شكل سحابة العلامة. للتحقق من صحة SoftCloud، وإثبات نقاط قوتها، أجريت التجارب على وثائق البرنامج المختلفة. أظهرت النتائج قدرة SoftCloud على استرداد جميع العلامات وتردداتها بشكل صحيح من وثائق البرنامج المتاحة.

**الكلمات الدالة:** هندسة البرمجيات، تصور البرمجيات، وثائق البرنامج، سحابة العلامات.

## 1. Introduction

Tag cloud has become a widespread visualization and navigation technique in the software engineering domain (Emerson, 2014; Lohmann et al., 2009). Software artifacts visualization helps software developers to manage the complexity and size of the software system (Al-Msie'deen, 2019c). This study suggests a new approach called *SoftCloud* to visualize software artifacts as tag clouds. In general, the tag cloud is a visualization technique for the content of a particular document (Al-Msie'deen, 2019a). Tag cloud uses the font size to denote how often a particular tag has been repeated through documents, while the tag color is for decoration purposes only (Al-Msie'deen, 2019b).

Each tag in the cloud usually represents a single word, and tag importance has shown appropriate font color and size (Rinaldi, 2019). Most current studies use the static tag clouds to represent tags of the textual documents and web pages (Hearst and Rosner, 2008; Cui et al., 2010; García-Castro et al., 2009; Greene and Fischer, 2015). Current approaches that build the tag cloud from the software code are either incomplete (*i.e.,* use either classes or methods) or do not perform pre-processing of the tag before adding it to the cloud (such as returning the English word to its root) (Emerson, 2014; Emerson et al., 2013a; Emerson et al., 2013b; Deaker et al., 2011; Cottrell et al., 2009; Anslow et al., 2008; Stocker, 2011; Martinez et al., 2016; Bajracharya et al., 2010). The literature has shown very limited work to mine tag cloud using different software artifacts (*cf.* Section 2). Figure 1 displays an example of a tag cloud — *SoftCloud's* representation of the abstract text of this paper.

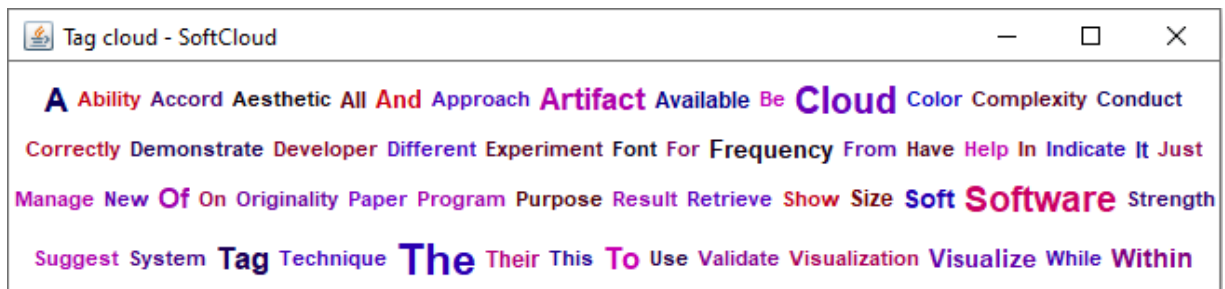

**Figure 1.** Tag cloud summarizing the abstract of this paper.

In this work, the software *artifacts* are any documents related to the software system. This paper considers any document resulting from the software development process as an artifact. Thus, the following documents are artifacts of the software: source code, commented code (*i.e.*, Javadoc), design documents such as software architecture document (Rational software corp., 2001), and so on. Javadoc is a software artifact developed by software experts to summarize the software code (Kramer, 1999). SoftCloud considers all software identifier names (*i.e.*, package, class, method, and attribute names) inside the code artifact.

In this paper, tag cloud displays the most common tags across software artifacts. In the tag cloud, some tags appear in different font sizes. However, some tags appear important more than other tags. The number of times a tag repeats within a software artifact determines the font size of this tag in the cloud (Yonezawa et al., 2020). However, this allows the software developer to see the most common tags as well as the unique tags in the tag cloud.

SoftCloud accepts any software artifacts as input. However, based on its parser, SoftCloud extracts all software artifact words. After that, it divides words into their constituent words. Then, it obtains the word roots. Then, it determines the weight of each tag based on its frequency across software artifacts. After that, it arranges tags in standard form. Tags are arranged according to their frequency, random or alphabetical. Finally, SoftCloud produces the tag clouds as outputs (*cf.* Figure 2).

SoftCloud is detailed in the rest of this paper as follows. Section 2 discusses the related work. Section 3 describes the SoftCloud approach step-by-step. Section 4 presents the experiments were conducted to validate SoftCloud's approach. Finally, section 5 concludes and provides future work of SoftCloud.

## 2. Related Work and Comparison with SoftCloud

This section presents the related work related to SoftCloud contributions. It also gives a concise summary of the diverse approaches and shows the need of suggesting SoftCloud's approach.

In the software engineering field, industrial tools and academic research have not focused on tag clouds as a popular visualization technique. Few studies have proposed the idea of visualizing the software artifacts as a tag cloud (Emerson, 2014; Emerson et al., 2013a; Emerson et al., 2013b; Deaker et al., 2011; Cottrell et al., 2009; Anslow et al., 2008; Stocker, 2011; Martinez et al., 2016; Bajracharya et al., 2010).

This section is limited to providing works very close to the contributions of SoftCloud. In the related work, each approach receives one type of software artifact as input. There is no generic approach to dealing with different software artifacts. Some existing works deal only with one artifact, such as software code or Javadoc (Al-Msie'deen, 2019b; Al-Msie'deen, 2019c). The approach proposed in this study used different software artifacts as inputs. Besides, SoftCloud listing some user tasks on the tag cloud, such as: finding a particular tag, and finding the most common tags, and so on.

Anslow *et al.* (Anslow et al., 2008) used a tag cloud to visualize software classes. Cottrell *et al.* (Cottrell et al., 2009) proposed an approach to visualize software methods as tag clouds. Sourcecloud (Stocker, 2011) created a tag cloud for software classes. Al-Msie'deen (Al-Msie'deen, 2019c) used a tag cloud to visualize software source code, while, Al-Msie'deen (Al-Msie'deen, 2019b) visualized JavaDocs file as a tag cloud. Also, a tag cloud is used in the *Sourcerer* API Search (Bajracharya et al., 2010) to visualize the code repository. Table 1 presents a comparison between the selected tag cloud studies (*i.e.*, small survey). The author evaluates the studied approaches according to the following criteria: inputs, outputs, cloud layout, and tag order.

**Table 1.** Selected main studies related to SoftCloud.

| ID | Inputs | | | | | | | | Outputs | | | Layout | | Order | | |
|---|---|---|---|---|---|---|---|---|---|---|---|---|---|---|---|---|
| Cf. Paper # | Packages | Classes | Attributes | Methods | JavaDocs | Code repositories | Specific text | Code blocks | Tag cloud | Block names | Code labels | Typewriter | Spiral | Alphabetical | Random | Frequency |
| 1 | ✵ | ✵ | ✵ | ✵ | | | | | ✦ | | | ✇ | | ✵ | | |
| 2 | | | | | ✵ | | | | ✦ | | | ✇ | | ✵ | | |
| 3 | | ✵ | | | | | | | ✦ | | | ✇ | | ✵ | | |
| 4 | | ✵ | | ✵ | | | | | ✦ | | | ✇ | ✇ | ✵ | | |
| 5 | | | | ✵ | | | | | ✦ | | | ✇ | | ✵ | | |
| 6 | | ✵ | | | | | | | ✦ | | | ✇ | | ✵ | | |
| 7 | | | | | | | | ✵ | | ✦ | | ✇ | | ✵ | | |
| 8 | | | | ✵ | | | | | ✦ | | | ✇ | ✇ | ✵ | ✵ | |
| 9 | ✵ | ✵ | ✵ | ✵ | | | | | | | ✦ | ✇ | | ✵ | | |

| 10 |  |  |  |  | ⚘ |  |  | ◆ |  |  | ⚘ |  | ⚘ |  |  |
| 11 |  |  |  |  |  | ⚘ |  | ◆ |  |  | ⚘ |  | ⚘ |  |  |
| 12 | ⚘ | ⚘ | ⚘ | ⚘ | ⚘ |  |  | ◆ |  |  | ⚘ | ⚘ | ⚘ | ⚘ | ⚘ |

| Paper # | Author(s) | Publication type |
|---|---|---|
| 1 | Al-Msie'deen, 2019c | Journal |
| 2 | Al-Msie'deen, 2019b | Journal |
| 3 | Anslow et al., 2008 | Conference |
| 4 | Deaker et al., 2011 | Technical report |
| 5 | Cottrell et al., 2009 | IEEE International Workshop |
| 6 | Stocker, 2011 | Eclipse plug-in |
| 7 | Martinez et al., 2016 | Conference |
| 8 | Emerson, 2014; Emerson et al., 2013ab | MSc Thesis, Conferences |
| 9 | Al-Msie'deen, 2018 | Journal |
| 10 | Bajracharya et al., 2010 | Conference |
| 11 | Feinberg, 2013 | Tool |
| 12 | SoftCloud | Journal |

The brief overview of the current approaches shows the need to suggest an approach to visualize different software artifacts as a tag cloud. SoftCloud's approach deals with different software artifacts such as source code, design documents, and JavaDocs. On the other hand, SoftCloud's approach performs preprocessing of the tag before adding it to the cloud, where it separates the words based on the camel-case splitting method, and then returns each word to its origin. Also, SoftCloud introduces some useful filters and user tasks (*e.g.*, search tasks) within the cloud.

## 3. SoftCloud Step by Step

This section gives an overview of SoftCloud approach and describes the approach step-by-step.

The study presented in this paper exploits the *tag cloud visualization technique* and applies it to the software engineering domain. The *originality* of this approach is that it receives as inputs different software artifacts. Then, this approach generates the tag clouds to render the input information. SoftCloud's approach is designed to deal with the software engineering datasets challenges (*e.g.*, scale and complexity of software) using suitable visual mappings existing in tag clouds to render the dataset data.

To visualize a software dataset as a tag cloud, it is important to define *visual characteristics* that might influence perception within tag cloud such as cloud layout (*e.g.*, typewriter), tag order (*e.g.*, random), tag length (*e.g.*, a variable number of letters or an equal number of letters), tag position, font size (also font family, style, size, and color), and cloud background color. In addition, it is important to choose visual characteristics that are suitable for data mapping, such as font size.

Dataset needs the necessary pre-processing procedure to prepare it. In the approach proposed in this paper, pre-processing is carried out by extracting the words of the available software artifacts. Then, the words have divided into their constituent words, and then each word is returned to its original. In conclusion, word repetition has counted throughout software artifacts, and at last, tags are arranged through the cloud using a specific order. An overview of SoftCloud's approach is shown in Figure 2.

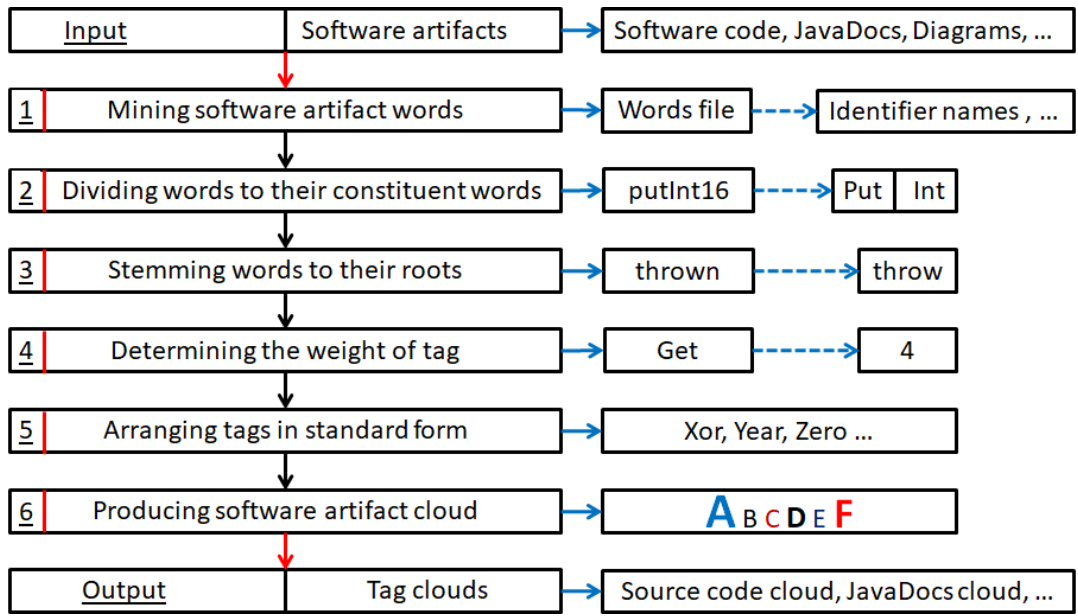

**Figure 2.** SoftCloud approach overview.

A tag cloud is a type of weighted list to visualize software artifact data (Jin, 2017), which gains growing attention and extra application opportunities in the software engineering field. As a demonstrative example, SoftCloud considers the source code of the *Rhino* software (Mozilla, 2012) and JavaDoc of *NanoXML* software (Scheemaecker, 2020). Rhino is an open-source application of JavaScript written completely in Java language. It is embedded in Java implementation to deliver scripting to end-users. J2SE 6 is used Rhino as the default Java scripting engine. NanoXML application is Java software for parsing XML documents. SoftCloud produces the artifact cloud in six phases are detailed below.

## 3.1. Mining Software Artifact Words

SoftCloud accepts the software artifact as input. Then, SoftCloud generates a words file as output. The words file contains all the words for the software artifact. Table 2 presents samples of words file contents of Rhino and NanoXML artifacts.

**Table 2.** Samples of words file contents.

| Software artifact | |
|---|---|
| **Rhino code** | **NanoXML JavaDoc** |
| org.mozilla.classfile | XMLParseException |
| itsExceptionTableTop | class |
| getClassName | summary |
| addLoadConstant | package |
| emptySubString | nanoxml |

SoftCloud considers the textual datasets (or words file), where the ideal datasets contain textual identifiers such as method names. The kind of dataset that would be ideal to show in a tag cloud is one that contains considerable amounts of textual information. Several datasets have contained this kind of information, in the form of identifiers, words, or labels. Software engineering datasets contain this type of data like package names and JavaDoc words.

## 3.2. Dividing Words to Their Constituent Words

SoftCloud divides the words extracted from the program's artifact into their constituent words. SoftCloud uses the *camel-case* splitting method to split artifact's words based on capital letters (*e.g.*, A-Z), special characters (*e.g.*, underscore), and numbers (*e.g.*, 0-9). Each word is divided into words based on the camel-case rules (Al-Msie'deen et al., 2014b).

**Table 3.** Samples showing examples of dividing words using camel-case.

| NanoXML JavaDoc words | | | | | Rhino identifier names | | |
|---|---|---|---|---|---|---|---|
| JavaDoc word | words | | | | Identifier name | Words | |
| | word1 | word2 | word3 | word4 | | word1 | word2 |
| NanoXML | nano | x | m | l | org.mozilla | org | mozilla |
| ParseException | parse | exception | | | itsFlags | its | flags |
| getLocalizedMessage | get | localized | message | | addField | add | field |
| printStackTrace | print | stack | trace | | putInt16 | put | int |
| getLineNr | get | line | nr | | unHex | un | hex |
| fillInStackTrace | fill | in | stack | trace | find_split | find | split |

The Camel-case method is easy and widely used for dividing words (Al-Msie'deen et al., 2014a). For instance: *getMaximumInterpreterStackDepth* identifier name has split into *get*, *maximum*, *interpreter*, *stack*, and *depth*. Table 3 presents samples of word splitting from Rhino and NanoXML software.

### 3.3. Stemming Words to Their Roots

*Stemming* is the text normalization (or called word normalization) technique, in the field of software engineering word normalization is used to prepare words for more processing. Stemming is a way of stripping attaches from words to form the word root (*e.g.*, protected to protect). The word root generated by SoftCloud does not have to be the real word itself. Stemmer is used in SoftCloud to return the word to its word root. In SoftCloud, stemming was performed through *WordNet* (Fellbaum, 1998). SoftCloud relies on WordNet dictionary to swap English words with their roots or stems (Princeton university, 2010).

**Table 4.** Examples of returning English words to their roots or origins.

| Rhino code words | | NanoXML JavaDoc words | |
|---|---|---|---|
| Identifier word | Root or stem | JavaDoc word | Root or stem |
| synchronized | synchronize | indicates | indicate |
| interfaces | interface | extends | extend |
| reserved | reserve | thrown | throw |
| parameters | parameter | parsing | parse |
| arguments | argument | occurred | occur |

In SoftCloud, stemming is a method of changing an artifact word to its root. The word root is the final form of the word that will appear in the cloud as a tag. SoftCloud stemmer accepts as an input English word and generates as output word root (or tag). For instance, the words *parsing*, *parses*, and *parsed* all have the same root/stem which is *parse*. Sometimes, the WordNet may not be dependable in all cases to return word root. In this case, SoftCloud returns the word itself as being the root of the word. Table 4 shows examples of the word stems from Rhino and NanoXML software artifacts.

### 3.4. Determining the Weight of Tag

In SoftCloud, tag weight gives a sign about tag frequency across software artifact words. In this stage, a weight is assigned to each tag based on its occurrences in software artifact words. Table 5 displays examples of tags and their weights from Rhino and NanoXML software artifacts.

**Table 5.** Examples of tags and their weights from Rhino and NanoXML artifacts.

| Rhino code tags | | NanoXML JavaDoc tags | |
|---|---|---|---|
| Tag | Weight | Tag | Weight |
| Activation | 20 | Exception | 10 |
| Adapter | 24 | From | 2 |
| Add | 134 | Get | 4 |
| And | 35 | Java | 7 |
| Arg | 12 | Line | 6 |

In fact, the number of times a tag is repeated is a very important indication of the importance of this tag in the software artifact. For instance, in drawing shapes software (Al-Msie'deen, 2019c), the *shape* tag arose *thirteen* times across software source code, so the given weight of this tag is *thirteen*. The font size for the tag in the mined cloud is the number of times the tag is repeated throughout the software artifact document. Tags that appear with a large font size are more important than others.

### 3.5. Arranging Tags in Standard Form

SoftCloud uses typewriter-style to arrange tags in the cloud from left to right and from top to bottom. SoftCloud displays tags in the cloud in alphabetical order (*i.e.*, a-z). Software developer looks more able to find tags in alphabetically ordered clouds (Al-Msie'deen, 2019c). Table 6 shows examples of tags in alphabetical order.

**Table 6.** Examples of tags in alphabetical order.

| Rhino code tags | | NanoXML JavaDoc tags | |
|---|---|---|---|
| Unordered tags | Tags in alphabetical order | Unordered tags | Tags in alphabetical order |
| Mozilla | Xmlend | Nano | Or |
| Classfile | Xop | X | Package |
| Class | Xor | M | Parse |
| File | Year | L | Print |
| Writer | Yield | Class | Public |
| Acc | Z | Parse | Runtime |
| Public | Zero | Exception | Stack |

On the other hand, SoftCloud allows the developer to arrange the tags according to their *frequency*. Tags have arranged within the cloud from the highest to lowest frequency. If some tags are equal in frequency, then SoftCloud sorts these tags alphabetically. Figure 3 shows the generated tag cloud after applying the *frequency order* filter.

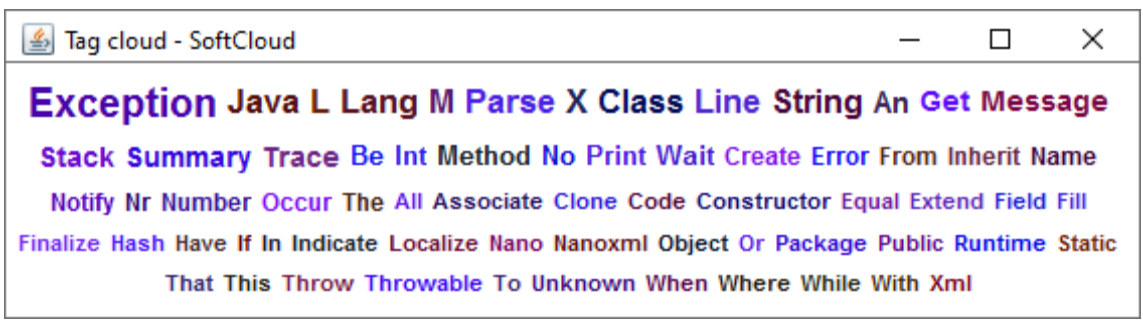

**Figure 3.** A tag cloud produced from JavaDoc of XMLParseException class of NanoXML.

In this cloud, tags appear according to their importance. The most important tags appear first in the cloud. The tag cloud in Figure 3 shows that the most common tag in the JavaDoc of XMLParseException class is an *exception*. The most common tags have been displayed in larger fonts.

### 3.6. Producing Software Artifact Cloud

In SoftCloud, the dataset is extracted first from the software artifact. Then the dataset words are divided into their constituent words. After that, each word is returned to its root. Later, the weights are determined for the tags, and then the software engineer determines the appropriate arrangement of the tags in the cloud. Finally, the cloud has been created. As an example, SoftCloud uses the JavaDoc for XMLParseException class of NanoXML software. Table 7 shows JavaDoc of XMLParseException class.

**Table 7.** JavaDoc of XMLParseException class of NanoXML (Scheemaecker, 2020).

| Class Summary |
|---|
| Package nanoxml.XMLParseException |
| public class XMLParseException |
| extends java.lang.RuntimeException |
| An XMLParseException is thrown when an error occurs while parsing an Xml string. |

| **Field Summary** | |
|---|---|
| static int | No_Line, indicates that no line number has been associated with this exception. |

| Constructor Summary |
|---|
| XMLParseException(java.lang.String name, int lineNr, java.lang.String message), creates an exception. |
| XMLParseException(java.lang.String name, java.lang.String message), creates an exception. |

| **Method Summary** | |
|---|---|
| int | getLineNr(), Where the error occurred, or No_Line if the line number is unknown. |

| Methods inherited from class java.lang.Throwable |
|---|
| fillInStackTrace, getLocalizedMessage, getMessage, printStackTrace, printStackTrace, printStackTrace, toString |

| Methods inherited from class java.lang.Object |
|---|
| clone, equals, finalize, getClass, hashCode, notify, notifyAll, wait, wait, wait |

Figure 4 shows a tag cloud extracted from JavaDoc of XMLParseException class of NanoXML (*cf*. Table 7). This cloud contains all tags of JavaDoc. The number next to each tag is an indication of how often that tag is repeated within the software artifact. The mined tag cloud shows the rarest tags such as *when* and *unknown*.

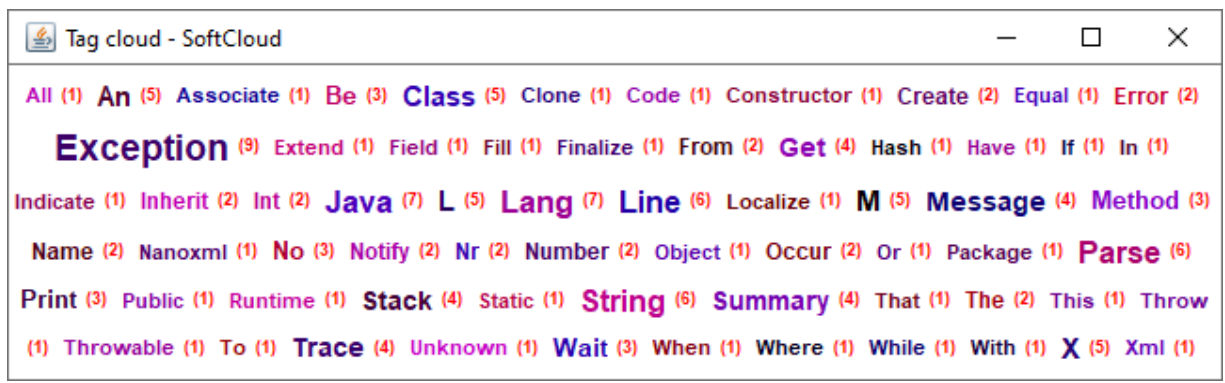

**Figure 4.** A tag cloud generated from JavaDoc of XMLParseException class.

SoftCloud contains several features to allow data exploration such as filtering data and handling large scale data. These features are the most important to software engineering datasets. SoftCloud prototype is formed to extract tag clouds from different software artifacts. SoftCloud prototype is available at author page (Al-Msie'deen, 2021a).

## 4. Experimentation

This section presents the experiments conducted in this research to display its soundness and presents different software artifacts. Also, it shows the obtained results for some artifacts and presenting the threats to the validity of SoftCloud. Figure 5 shows mined tag cloud from Rhino software. SoftCloud algorithms need *22697 ms* to generate tag cloud from Rhino artifact. The most common tags (*resp*. the rarest tags) across Rhino artifacts are presented in Table 8.

**Table 8.** Tags mined from Rhino artifact.

| The most common tags | | The rarest tags | |
|---|---|---|---|
| **Tag** | **Frequency** | **Tag** | **Frequency** |
| Get | 510 | Zone | 2 |
| Id | 444 | Collect | 1 |
| Set | 172 | W | 4 |
| Name | 168 | After | 3 |
| Class | 159 | Yield | 6 |
| The *number of tags* across Rhino code is equal to *1095*. | | | |
| The *execution time* of SoftCloud in *ms* is equal to *22697*. | | | |

The success of a SoftCloud is measured by three metrics: *precision*, *recall*, and *F-Measure* (Al-Msie'deen, 2019b). Precision and recall give a value of one, if the tag and its frequency in the cloud are the same as tag frequency in the software artifact. F-Measure gives a value of one in cases where both precision and recall are one (Al-Msie'deen, 2014). SoftCloud evaluation metrics have values between zero and one.

For a specific tag within the cloud, a *precision* metric is a percentage of correctly retrieved tag frequencies to the total number of retrieved tag frequencies (*cf*. equation in Table 9), whereas *recall* metric is the percentage of correctly retrieved tag frequencies to the total number of relevant tag frequencies. The *F-Measure* metric combines recall and precision in one value (Al-Msie'deen, 2014). An example of the calculation of these three metrics are presented in Table 9.

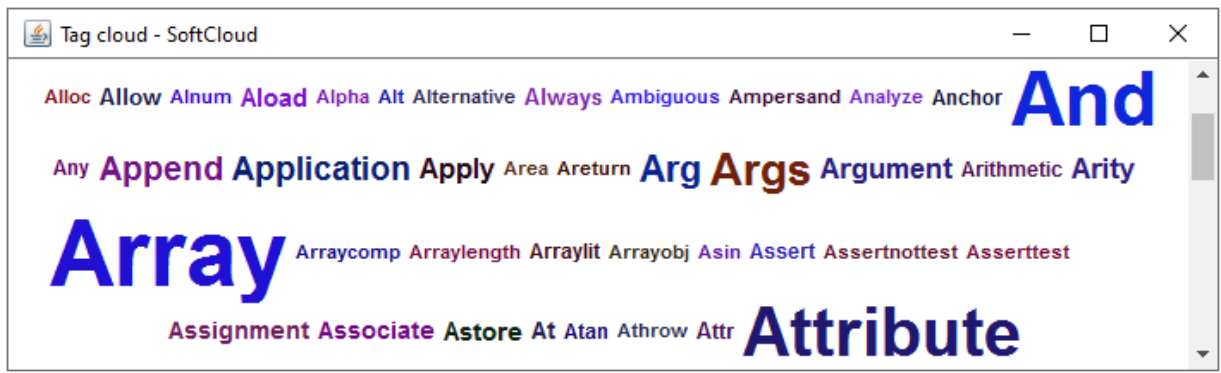

**Figure 5.** A tag cloud generated from Rhino artifact.

An illustrative example is introduced in Table 9 to show: 1) how to calculate these measures for a *trace* tag from JavaDoc of XMLParseException class (*cf*. Table 7), and 2) the equation of each measure. Moreover, 3) how to compute these measures based on some samples (not related to SoftCloud experimentation).

**Table 9.** Standard SoftCloud evaluation metrics: precision, recall, and F-Measure.

| Tag | relevant tag frequency | correctly retrieved tag frequencies | retrieved tag frequencies |
|---|---|---|---|
| **Trace** | 4 | 4 | 4 |
| **Metric** | **Precision** | **Recall** | **F−Measure** |
| **Value** | 1 | 1 | 1 |
| **Precision** = \|{relevant tag frequencies} ∩ {retrieved tag frequencies}\| / \|{retrieved tag frequencies}\| | | | |
| **Recall** = \|{relevant tag frequencies} ∩ {retrieved tag frequencies}\| / \|{relevant tag frequencies }\| | | | |
| **F−Measure** = 2 × [(Precision × Recall) / (Precision + Recall)] | | | |
| Tag | relevant tag frequency | correctly retrieved tag frequencies | retrieved tag frequencies |
| **Notify** | 100 | 50 | 150 |
| **Metric** | **Precision** | **Recall** | **F−Measure** |
| **Value** | 0.3 | 0.5 | 0.4 |
| Tag | relevant tag frequency | correctly retrieved tag frequencies | retrieved tag frequencies |
| **Wait** | 70 | 70 | 100 |
| **Metric** | **Precision** | **Recall** | **F−Measure** |
| **Value** | 0.7 | 1 | 0.8 |

Low precision (*e.g.*, precision = 0.1) leads to low trust in the proposed system (*i.e.*, too much noise). On the other hand, low recall (*e.g.*, recall = 0.1) leads to unawareness and inefficiency of the suggested approach (*i.e.*, too many missing frequencies for the tag). Table 10 summarizes the obtained results of some tags from Rhino and NanoXML software artifacts.

**Table 10.** Tags mined from Rhino and NanoXML software artifacts.

| Software | Tag | Tag within the cloud | Tag within the artifact | SoftCloud evaluation metrics | | |
|---|---|---|---|---|---|---|
| | | | | Precision | Recall | F-Measure |
| Rhino | A | 37 | 37 | 1 | 1 | 1 |
| | And | 35 | 35 | 1 | 1 | 1 |
| | Arg | 12 | 12 | 1 | 1 | 1 |
| NanoXML | Get | 4 | 4 | 1 | 1 | 1 |
| | X | 7 | 7 | 1 | 1 | 1 |
| | An | 5 | 5 | 1 | 1 | 1 |

Results display that precision value is one of all mined tags. Thus, all frequencies of the retrieved tag are relevant. Recall metric value equals one of all mined tags. Hence, all relevant tag frequencies are retrieved. F-Measure value equals one of all mined tags. Consequently, all relevant tag frequencies are recovered, and only the relevant tag frequencies are recovered. The results demonstrate the efficiency and ability of SoftCloud to accurately retrieve the correct frequency of tags from software artifacts. Figure 6 shows the tag cloud generated from the source code *summarization* of the *draw* method from *drawing shapes* software (Al-Msie'deen and Blasi, 2019).

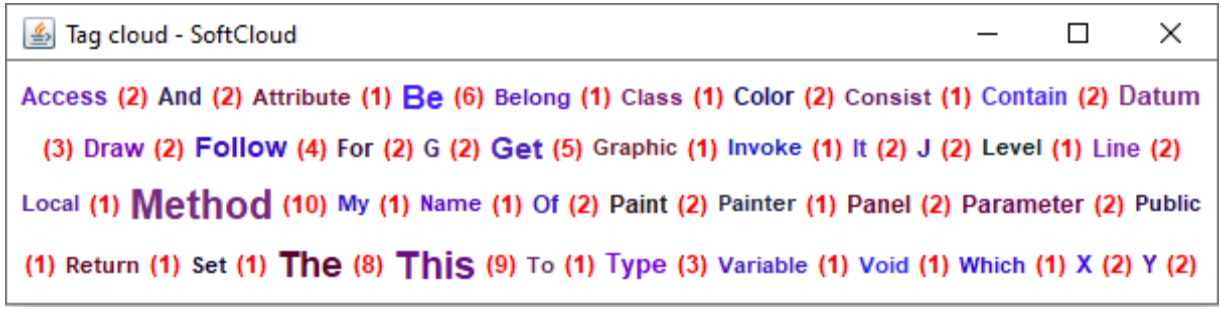

**Figure 6.** Tag cloud generated from source code summarization.

I have implemented numerous tag cloud layouts. Tags are positioned one at a time within the cloud, with the chosen order (*e.g.*, alphabetical order). SoftCloud layouts are typewriter and spiral layout. In typewriter layout tags are positioned left to right, jumping to a new line once the next tag cannot be positioned on the existing line. While, in a spiral layout the first tag is positioned in the middle of the cloud, with consecutive tags are being positioned around it in a spiral style. Figure 7 expressions the same data set in Figure 6 with a spiral layout chosen. This layout is less appropriate for some tasks, including discovering a particular tag or emphasizing its absence.

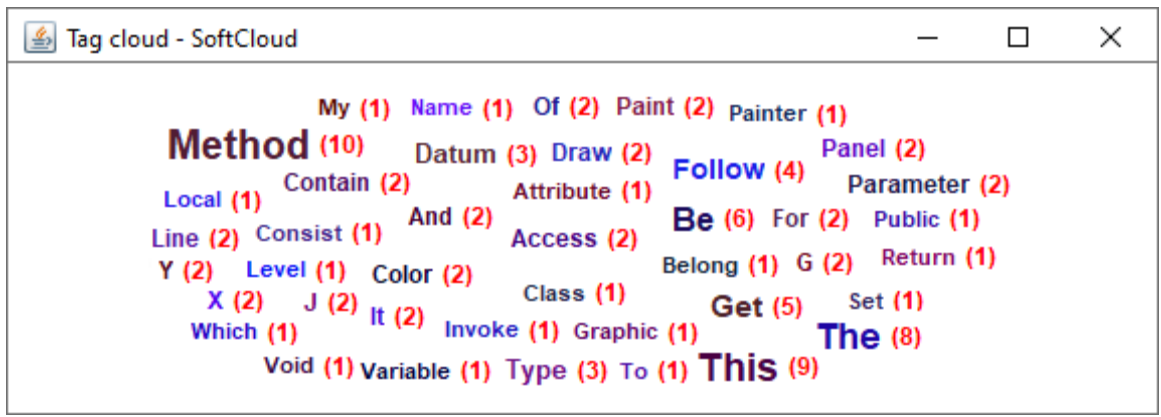

**Figure 7.** Tag cloud with spiral layout.

Figure 8 shows the tag cloud generated from user and system requirement of *registration* service. This requirement is included in the *requirements document* of the interactive multimedia magazine application (Al-Msie'deen, 2021b). A software engineer can extract the tag cloud from several software artifacts such as use-case description (Al-Msie'deen, 2008), use-case diagram (Alfrijat and Al-Msie'deen, 2010), software identifiers map (Al-Msie'deen and Blasi, 2021), and feature descriptions (Salman et al., 2012).

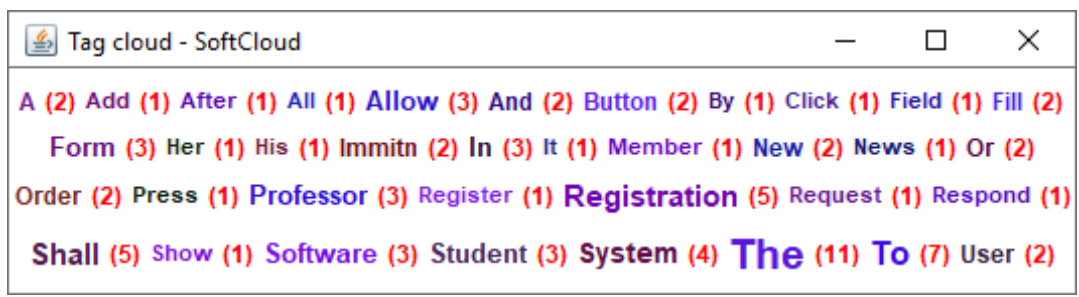

**Figure 8.** Tag cloud generated from the software requirements specification document.

The software *architecture document* is one of the software's artifacts. This document is a design document. Figure 9 represents the tag cloud generated from a software *architecture document* of the collegiate sports paging system (Rational software corp., 2001).

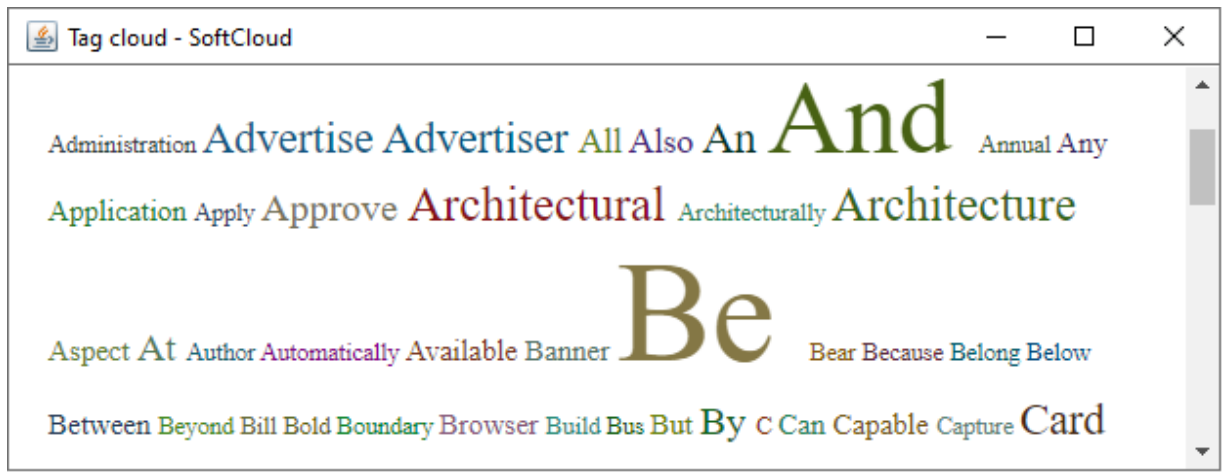

**Figure 9.** Tag cloud summarizing architecture document of collegiate sports paging system.

The user of SoftCloud has the choice of filtering the tag text to a *fixed number* of letters (*e.g.,* 10 letters). This filter has two aims (*cf.* Figure 10), to exploit available space in the cloud, and to minimize any side effect a larger number of letters in a tag may have on user awareness (*i.e.,* eye attention).

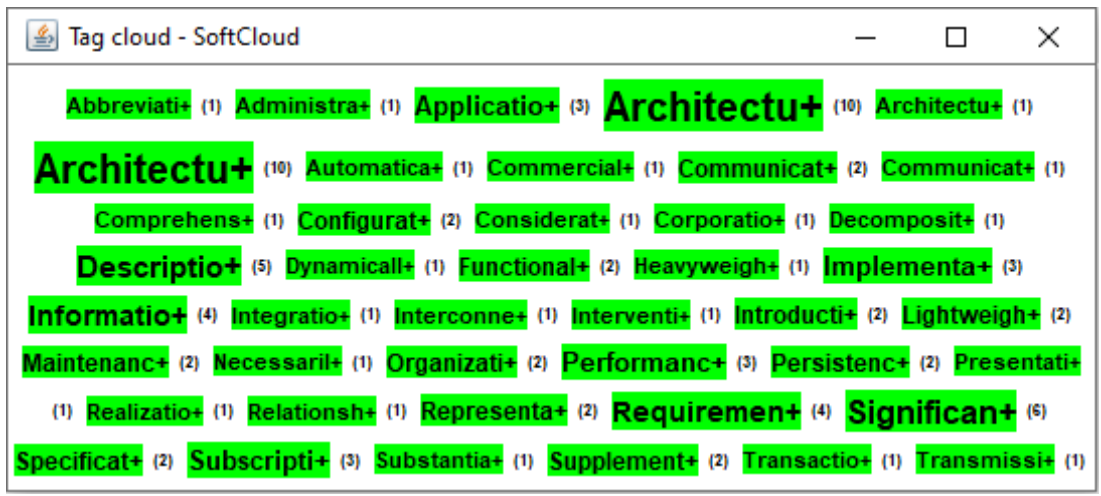

**Figure 10.** A tag cloud generated by using fixed number filter.

The *threat to the validity* of SoftCloud is that the existing prototype considers only a Java code artifact. Moreover, when a software engineer uses mixture words inside software artifacts (*e.g.*, SeTStandardS) the camel-case splitting method cannot deal with it (or should be enhanced with other methods). The WordNet dictionary may not be dependable in all cases to reveal the word root. Currently, SoftCloud is missing some filters, for instance, it does not filter tag names that are textually similar.

## 5. Conclusion and Future Directions of SoftCloud

This paper proposed a new approach to visualize software artifacts as a tag cloud. However, SoftCloud has executed on different software artifacts. Such as rhino, nanoXML, drawing shapes, interactive multimedia magazines, and collegiate sports paging software artifacts. The results were showed that all tags and their occurrences are mined correctly from software artifacts. However, the mined tag clouds have shown the most common and rarest tags. Also, Tags have been arranged randomly, alphabetically, or according to their frequency. Also, Tags within the cloud are filtered based on their frequency or length. Besides, Clouds have been organized according to a typewriter or spiral layout. For future work, some user tasks will be added to the cloud and use new cloud layouts. Finally, there is an urgent need for a comprehensive survey providing all studies related to the tag cloud techniques in the software engineering domain.

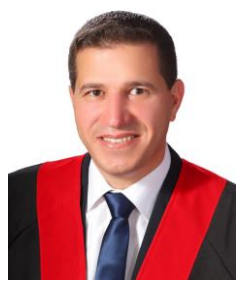

**Ra'Fat Al-Msie'Deen** is an Associate Professor at Mutah University since 2014. He received his PhD in Software Engineering from the University of Montpellier 2, Montpellier – France, in 2014. He received his MSc in Information Technology from the University Utara Malaysia, Kedah – Malaysia, in 2009. He got his BSc in Computer Science from Al-Hussein Bin Talal University, Ma'an – Jordan, in 2007. His research interests include software engineering, software product line engineering, and formal concept analysis.